\documentclass[aps,preprint,nofootinbib,superscriptaddress]{revtex4}

\usepackage{mathrsfs}
\usepackage{amsmath}
\usepackage{array}
\usepackage{verbatim}
\usepackage{epsfig}
\usepackage{epsf}
\usepackage{CJK}
\usepackage{graphicx}
\begin{document}

\title{Mass and isospin dependence of symmetry energy coefficients of finite nuclei}

\author{Ning Wang}
\affiliation{Department of Physics, Guangxi Normal University,
Guilin 541004, P. R. China}

\author{Min Liu}
\email{lium_816@hotmail.com} \affiliation{Department of Physics,
Guangxi Normal University, Guilin 541004, P. R. China}
\affiliation{ College of Nuclear Science and Technology, Beijing
Normal University, Beijing, 100875, P. R. China}

\begin{abstract}
Mass and isospin dependence of symmetry energy coefficients
$a_{\rm sym}$ of finite nuclei is investigated with the measured
nuclear masses incorporating the liquid drop mass formula. The
enhanced $a_{\rm sym}$ for nearly symmetric nuclei are observed.
To describe the mass and isospin dependence of $a_{\rm sym}$, a
modified formula based on the conventional surface-symmetry term
is proposed and the corresponding rms deviation of nuclear masses
is checked.
\end{abstract}
\maketitle

The study of nuclear symmetry energy has attracted great attention
in recent years both theoretically and experimentally
\cite{Dani,Sam,Kirson,Mend,Kol}. The mass dependence of symmetry
energy coefficients has been clearly observed and the obtained
symmetry energy coefficients of finite nuclei are considerably
smaller than the symmetry energy coefficient of nuclear matter at
saturation density. The symmetry energy coefficient of finite
nuclei is usually extracted by directly fitting the measured
nuclear masses with different versions of the liquid drop mass
formula. Some different forms for describing the mass dependence
of symmetry energy coefficients of finite nuclei, which divide the
symmetry energy of a nucleus into the volume and surface
contributions, have been proposed in Refs.
\cite{Mend,Dani,Kirson}. The volume symmetry energy of nucleus
corresponds to that of nuclear matter at saturation density. In
this work, we investigate the symmetry energy coefficients of
nuclei, especially the isospin dependence of the symmetry energy
coefficient, based on the more than 2000 precisely measured
nuclear masses \cite{Audi}.

We start with the well known liquid drop formula. The liquid drop
energy of a nucleus is described by a Bethe-Weizs\"acker mass
formula \cite{Bethe},
\begin{eqnarray}
E_{\rm LD}(A,Z)=a_{v} A + a_{s} A^{2/3}+ E_{\rm Coul} + a_{\rm
sym} I ^2 A,
\end{eqnarray}
neglecting the pairing term. Where, $I=(N-Z)/A$ denote the
isospin-asymmetry. The symmetry energy coefficient $a_{\rm sym}$
is conventionally expressed as a function of mass number $A$
\cite{Dani}. The Coulomb energy is written as
\begin{eqnarray}
E_{\rm Coul}=a_{c} \frac{Z(Z-1)}{A^{1/3}} \left ( 1- 0.76 Z^{-2/3}
\right)
\end{eqnarray}
with the coefficient $a_c = 0.71$ MeV. Inserting Eq.(2) into
Eq.(1) and using the relation $Z= \frac{A}{2}(1-I)$, the liquid
drop energy per particle $\varepsilon_{\rm LD} = E_{\rm LD}/A$ can
be expressed as a function of mass number $A$ and isospin
asymmetry $I$. Performing a partial derivative of
$\varepsilon_{\rm LD} (A,I)$ with respect to the isospin asymmetry
$I$, the symmetry energy coefficient can be expressed as
\begin{eqnarray}
a_{\rm sym} &=& a_{\rm sym}^{(0)}-\frac{I}{2} \frac{\partial
a_{\rm sym}}{\partial  I}  = a_{\rm sym}^{(0)}+ a_{\rm
sym}^{(1)}+...
\end{eqnarray}
with
\begin{eqnarray}
a_{\rm sym}^{(0)}=\left ( \frac {\partial \varepsilon_{\rm LD}
(A,I)}{\partial I}- \frac{1}{A} \frac{\partial E_ {\rm Coul}(A,I)
}{ \partial I} \right ) / (2I).
\end{eqnarray}
Omitting the microscopic shell and pairing corrections and the
corrections from nuclear deformation, the values of $\frac
{\partial \varepsilon_{\rm LD} (A,I)}{\partial I}$ could be
obtained from the measured energy per particle $\varepsilon_{\rm
exp}$ \cite{Audi},
\begin{eqnarray}
\frac {\partial \varepsilon_{\rm LD} (A,I)}{\partial I} \approx
\frac{\varepsilon_{\rm exp} (A,I_2)-\varepsilon_{\rm exp}
(A,I_1)}{I_2-I_1}.
\end{eqnarray}
Where, $I_1$ and $I_2$ denote the isospin asymmetry of nuclei
$(A,Z-\Delta Z)$ and $(A, Z+\Delta Z)$, respectively. We take
$\Delta Z=1$ in this work. Through iterations to $a_{\rm sym}$ in
Eq.(3),
\begin{eqnarray}
a_{\rm sym} = a_{\rm sym}^{(0)}- \frac{I}{2} \frac{\partial
}{\partial  I} \left [ a_{\rm sym}^{(0)}- \frac{I}{2}
\frac{\partial }{\partial  I} \left ( a_{\rm sym}^{(0)}-
\frac{I}{2} \frac{\partial }{\partial  I} [ a_{\rm sym}^{(0)} -...
] \right ) \right ],
\end{eqnarray}
one can obtain the expression of $a_{\rm sym}^{(1)}$ which
includes all terms with $\frac{\partial }{\partial  I}a_{\rm
sym}^{(0)} $ in Eq.(6),
\begin{eqnarray}
a_{\rm sym}^{(1)}= \sum_{n=1}^{\infty} \left ( - \frac{1}{2}
\right )^n I \frac{\partial a_{\rm sym}^{(0)}}{\partial I} =
-\frac{I}{3} \frac{\partial a_{\rm sym}^{(0)}}{\partial I}.
\end{eqnarray}

\begin{figure}
\includegraphics[angle=0,width=1.0 \textwidth]{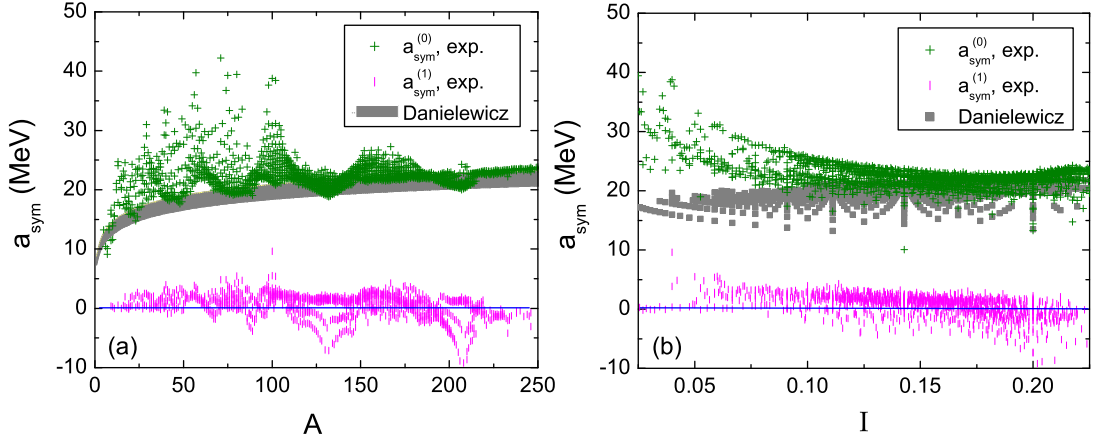}
\caption{(Color online) Symmetry-energy coefficients of nuclei as
a function of nuclear mass number (a) and of isospin asymmetry
(b). The shades denote the results of Danielewicz et al
\cite{Dani}. The crosses and the short dashes denote the extracted
$a_{\rm sym}^{(0)}$ and $a_{\rm sym}^{(1)}$ terms of the symmetry
energy coefficients from the measured nuclear masses,
respectively.}
\end{figure}

In Fig.1 (a), we show the extracted symmetry energy coefficients
of nuclei as a function of nuclear mass number. The crosses and
the short dashes denote the extracted $a_{\rm sym}^{(0)}$ and
$a_{\rm sym}^{(1)}$ terms of the symmetry energy coefficients from
the measured nuclear masses, respectively. One can see that the
contribution of $a_{\rm sym}^{(1)}$ term is much smaller than that
of $a_{\rm sym}^{(0)}$ term for most nuclei since $I$ is a small
quantity. The relatively large fluctuations in $a_{\rm sym}^{(1)}$
for heavy nuclei are mainly caused by the shell effects. The
shades denote the results of Danielewicz et al. \cite{Dani},
\begin{eqnarray}
a_{\rm sym}=c_{\rm sym} \left [ 1+ \kappa A^{-1/3} \right ]^{-1}.
\end{eqnarray}
The extracted $a_{\rm sym}^{(0)}$ term of the symmetry energy
coefficients for heavy nuclei are comparable to the results of
Danielewicz et al. For intermediate and light nuclei, there exists
obvious oscillations and fluctuations in the extracted $a_{\rm
sym}^{(0)}$ which are probably caused by the shell effects and
other nuclear structure effects. In the region $A<120$, the
extracted $a_{\rm sym}^{(0)}$ are generally higher than the
results of Danielewicz et al. In Fig.1(b), we show the same data
as in Fig.1(a), but as a function of isospin asymmetry $I$. One
can find that the obtained $a_{\rm sym}^{(0)}$ term of the
symmetry energy coefficients somewhat depend on the corresponding
isospin asymmetry of nuclei, especially for nearly symmetric
nuclei. The $a_{\rm sym}^{(0)}$ term of the symmetry energy
coefficient obviously increases with the decrease of asymmetry.
The dependence of symmetry energy coefficient on the asymmetry of
nucleus, especially that $a_{\rm sym}$ increases with increasing
proton fraction of the system is also found in \cite{Sam}. The
results from Danielewicz et al. cannot reproduce the observed
trend of isospin dependence well.

To describe the isospin and mass dependence of symmetry energy
coefficient of nucleus, we propose a modified formula,
\begin{eqnarray}
a_{\rm sym}=c_{\rm sym}\left [1-
\frac{\kappa}{A^{1/3}}+\frac{2-|I|}{2+|I|A} \right ],
\end{eqnarray}
based on the conventional surface-symmetry term of liquid drop
model, with a small correction term from isospin asymmetry. The
introduced correction term approximately describes the Wigner
effect \cite{Moll95,Lun03,Royer,Sat,Myers} of nuclei. The
introduced $I$ term in $a_{\rm sym}$ roughly leads to a correction
$E_W $ to the binding energy of the nucleus,
\begin{eqnarray}
E_W  = c_{\rm sym} I^2 A \left [\frac{2-|I|}{2+|I|A} \right ]
\approx 2 c_{\rm sym} |I|   - c_{\rm sym} |I|^2 +... ,
\end{eqnarray}
which is known as the Wigner term. In Fig.2, we show the Wigner
energies of nuclei calculated with different models. The open
circles and the crosses denotes the results in Ref.\cite{Royer}
and those of this work, respectively. The straight line denotes
the results of Satula et al. \cite{Sat}, i.e. $E_W \approx 47
|I|$. In \cite{Myers}, Myers and Swiatecki write the Wigner term
as $E_W = -C_0 \exp[-W|I|/C_0] \approx -C_0 + W|I|+...$, with
$C_0=10$ MeV, $W=42$ MeV. The results of this work are comparable
to those from Ref. \cite{Sat} for most nuclei.

\begin{figure}
\includegraphics[angle=0,width=0.6 \textwidth]{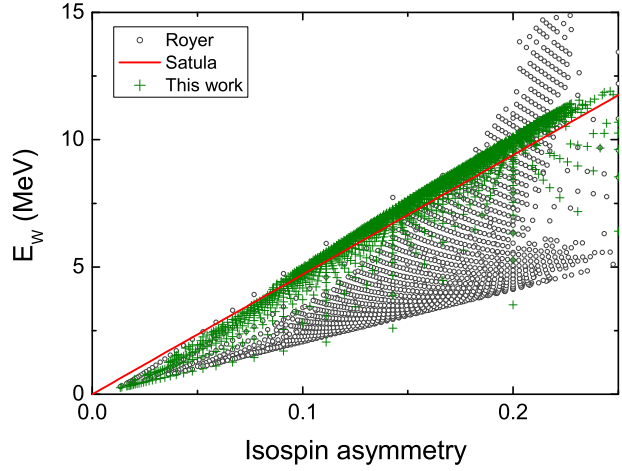}
\caption{(Color online) Wigner energies of nuclei calculated with
different models. The open circles and the straight line denote
the results in Ref. \cite{Royer} and those of Satula et al.
\cite{Sat}, respectively. The crosses denote results of this work,
with $c_{\rm sym} = 29.3$ MeV determined by fitting the 2149
measured nuclear masses. }
\end{figure}
\begin{figure}
\includegraphics[angle=0,width=0.95 \textwidth]{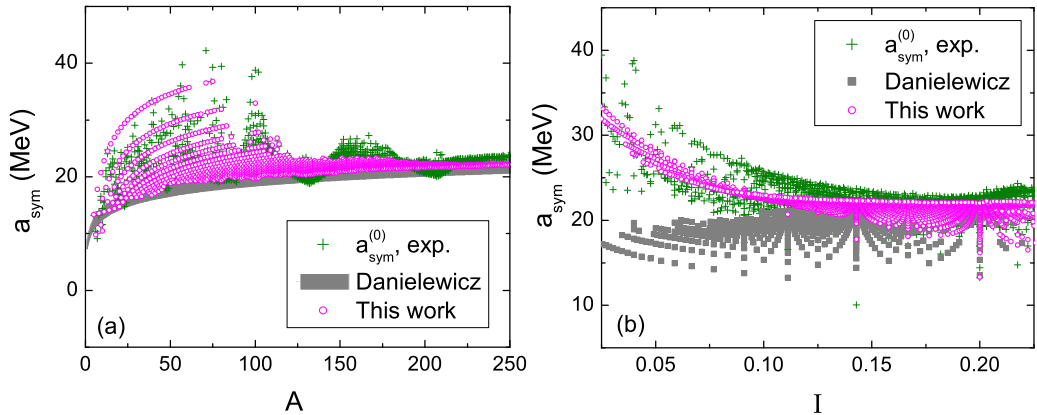}
\caption{(Color online) The same as Fig.1, but with the results of
Eq.(9) (open circles) for comparison.}
\end{figure}

With increasing of mass number $A$, the $a_{\rm sym}$ in Eq.(9)
has a finite value approaches $c_{\rm sym}$ which corresponds to
the symmetry energy coefficient of nuclear matter at saturation
density. The results from Eq.(9)  are shown in Fig.3 for
comparison, with $c_{\rm sym}=31$ MeV and $\kappa=2$ which are
obtained by fitting the extracted $a_{\rm sym}^{(0)}$. One can see
that the extracted $a_{\rm sym}^{(0)}$ can be reproduced
reasonably well. Furthermore, we have checked the rms deviations
of 2149 masses of nuclei with $N$ and $Z\ge8$ from the measured
data defined as $ \sigma^{2}=\frac{1}{m}\sum \left ( M_{\rm
exp}^{(i)} - M_{\rm th}^{(i)} \right )^2$  by taking different
forms of symmetry energy coefficients mentioned above
incorporating the liquid drop mass formula of Eq.(1). The obtained
rms deviations and the corresponding parameters of the liquid drop
formula are listed in Table 1. Adopting the form in Eq.(8), we
obtain an rms deviation of 2.71 MeV. With Eq.(9) for the symmetry
energy coefficient, the rms deviation is reduced to 2.55 MeV.
Compared with the case without the $I$ term being taken into
account, the rms deviation is reduced by $6\%$ (see Table 1).
Incorporating the semi-empirical mass formula in \cite{Wang}, the
rms deviation of the 2149 masses of nuclei can be considerably
reduced, falling to 0.516 MeV. Furthermore, when the isospin
dependence of symmetry energy coefficient is taken into account,
the obtained optimal $c_{\rm sym}$ changes from 26.09 to 29.38 MeV
which is close to the calculated symmetry energy coefficient of
nuclear matter at saturation density from the Skyrme energy
density functional \cite{Wang}.
\begin{table}
\caption{ rms $\sigma$ deviations between 2149 measured data and
predictions of Eq.(1) with different $a_{\rm sym}$ forms, and the
corresponding optimal parameters of liquid drop formula. }
\begin{tabular}{ccccccc}
 \hline\hline
  $a_{\rm sym}$ form   & $a_v$ (MeV) & $a_s$ (MeV)  & $a_c$ (MeV) & $c_{\rm sym}$ (MeV) & $\kappa$ & $\sigma$ (MeV)\\
\hline
Danielewicz   &   $-15.55$  &  18.18    &   0.71    &   27.39      & 1.28     & 2.71 \\

$c_{\rm sym}[1-  \kappa A^{-1/3}]$     &   $-15.57$  &  18.25    &   0.71    &   26.09      & 0.80     & 2.72 \\
This work     &   $-15.56$  &  18.11    &   0.71    &   29.38      & 1.52     & 2.55 \\
\hline\hline
\end{tabular}
\end{table}

In summary, the mass and isospin dependence of symmetry energy
coefficients $a_{\rm sym}$ of finite nuclei has been investigated
with the measured nuclear masses incorporating  the liquid drop
formula. For heavy nuclei, the extracted $a_{\rm sym}^{(0)}$ term
of the symmetry energy coefficients are consistent with the
results of Danielewicz et al. For light and intermediate nuclei,
there exists oscillations and fluctuations in the extracted
$a_{\rm sym}^{(0)}$. The isospin dependence of symmetry energy
coefficients, especially the enhanced $a_{\rm sym}$ in nearly
symmetric nuclei, has been observed. To describe the mass and
isospin dependence of $a_{\rm sym}$, we propose a modified formula
based on the conventional surface-symmetry term, with which the
isospin dependence of $a_{\rm sym}$ can be described reasonably
well and the rms deviation of nuclear masses from the experimental
data can be effectively reduced.

\begin{center}
\textbf{ACKNOWLEDGEMENTS}
\end{center}
This work was supported by National Natural Science Foundation of
China, Nos 10875031 and 10847004.

\end{document}